\documentclass[
noshowpacs,
amsmath,
amssymb,
amsfonts,
twocolumn,
reprint,
floatfix,
longbibliography,
]{revtex4-1}

\usepackage{rotating}
\usepackage{amsmath}
\usepackage{bbm}
\usepackage{subfigure}
\usepackage{color}
\usepackage{psfrag}

\newcommand{\ket}[1]{\left| \, #1 \, \right\rangle}
\newcommand{\braket}[2]{\left\langle \, #1 \,|\, #2 \, \right\rangle}

\def\bea{\begin{eqnarray}} 
\def\eea{\end{eqnarray}} 
\def\ben{\begin{equation}} 
\def\een{\end{equation}} 
\def\benu{\begin{enumerate}} 
\def\enu{\end{enumerate}}

\newcommand{\abs}[1]{\left| #1 \right|} 
\newcommand*{\citen}[1]{%
  \begingroup
    \romannumeral-`\x 
    \setcitestyle{numbers}%
    \cite{#1}%
  \endgroup   
}

\begin{document}

\title{A quantum reactive scattering perspective on electronic nonadiabaticity}

\author{Yang Peng} 
\author{Luca M. Ghiringhelli}
\author{Heiko Appel}
\email[E-Mail: \ ]{\{peng,ghiringhelli,appel\}@fhi-berlin.mpg.de}
\affiliation{
Fritz-Haber-Institut der Max-Planck-Gesellschaft, Faradayweg 4-6, 
14195, Berlin, Germany}

\date{\today}

\begin{abstract}
Based on quantum reactive-scattering theory, we propose a method for 
studying the electronic nonadiabaticity in collision processes
involving electron-ion rearrangements.
We investigate the state-to-state transition probability for electron-ion
rearrangements with two comparable approaches. In the first approach 
the information of the electron is only contained in the ground-state 
Born-Oppenheimer potential-energy surface, which is the starting point of 
common reactive-scattering calculations. In the second approach, the 
electron is explicitly taken into account and included in the calculations
at the same level as the ions. Hence, the deviation in the 
results between the two approaches directly reflects the electronic 
nonadiabaticity during the collision process. 
To illustrate the method, we apply it to the well-known 
proton-transfer model of Shin and Metiu (one electron and three ions), 
generalized by us in order to allow for reactive scattering channels. 
It is shown that our explicit electron approach is able to
capture electronic nonadiabaticity 
and the renormalization of the reaction barrier near the 
classical turning points of the potential in nuclear configuration
space. In contrast, system properties near the equilibrium 
geometry of the asymptotic scattering channels are hardly affected by 
electronic nonadiabatic effects. 
We also present an analytical expression for the transition amplitude of 
the asymmetric proton-transfer model based on the direct evaluation of 
integrals over the involved Airy functions.
\end{abstract}


\date{\today}

\maketitle

\section{Introduction}
\label{sec:Intro}
The fundamental understanding of elementary chemical reactions is an 
important subject in chemical physics. The development of molecular-beam 
scattering techniques has made it possible to experimentally study detailed
state-to-state dynamics of gas phase reactions\cite{Mikosch2008,Mikosch2013,Guo2012}.  
On the other hand, the interest for developing reactive-scattering theories 
to describe chemical reactions arose much earlier, shortly after the discovery 
of quantum mechanics\cite{hirschfelder1936,*hulburt1943}, when it was realized 
that the Born-Oppenheimer (BO) approximation leads naturally to the concept of 
potential-energy surfaces (PES), that govern the motion of atoms during a chemical
reaction. The PES play such a crucial role as the potential governing the 
dynamics of the nuclei, that almost all reactive-scattering approaches, either 
classical\cite{Bonnet2013} or quantum\cite{Nyman2013}, are using PES data as 
initial input.  
For many cases, the ground-state BO PES is sufficient for the description of
scattering events\cite{Hu2006}, since the motion of the nuclei is typically much 
slower than that of the electrons so that electrons can be assumed to be 
effectively in the ground state.
In this respect, the notion of electronic nonadiabaticity is identified with 
the set of all those ingredients that are missed when assuming that the motion 
of atoms is governed by the lowest (ground-state) PES. 

It is not uncommon to find situations where nonadiabaticity plays an important 
role\cite{Butler1998,Yarkony1996,Chu2006, RBaer2006,tully2012}. Such examples encompass
reactions involving light ions, charge transfer and photochemical 
processes\cite{Sheps2010,Woerner2011,Garand2008}. 
To capture this type of nonadiabaticity, reactive-scattering treatments involving 
excited PESs were developed\cite{shin2000}, which allow to take into account the 
electronic excitations during scattering. In general, these approaches use as 
{\em a priori} inputs, besides several ground- and excited-states PESs, the 
nonadiabatic coupling terms\cite{Baer2002} between different surfaces. Such 
coupling terms can nowadays be directly calculated through e.g.~the 
linear-response formalism of time-dependent 
density-functional theory\cite{RBaer2002,tavernelli2009,tavernelli2010,hu2011}. However, 
the multi-PES scattering approach typically only involves a small number of 
surfaces (in most cases only the ground and first-excited PES are considered) 
and may not always contain all essential ingredients of electronic nonadiabaticity.

In the work presented here, instead of directly utilizing several electronic surfaces 
and non-adiabatic couplings as input, we choose a specific coordinate system convenient 
for scattering calculations which allows us to quantify electronic nonadiabaticity from 
a perspective that differs from the normal treatment in the literature. As an illustration 
of our idea, and in order to make it as clear as possible, we restrict ourselves to a 
simple 1D collinear model, which was originally designed to study the nonadiabatic 
effects in a proton-transfer reaction\cite{shin1995} by Shin and Metiu. We emphasize, 
that our approach also remains valid for ab initio
Hamiltonians in full dimensionality. As many other recent studies on electronic nonadiabaticity in the 
literature\cite{Tolstikhin2013,Abedi2013,Vindel2013,Falge2012,hader2012,falge2011,Hader2011},
we solely restrict our discussion to a 1D model in order to simplify the mathematical
expressions and to highlight the essence of the underlying physics.

Our paper is organized as follows. In section \ref{sec:model},
we introduce the model used for the illustration of our scattering treatments 
on electronic nonadiabaticity. These treatments are discussed in detail
in section \ref{sec:scattering}, where we consider two approaches 
which can be compared in parallel. In the first approach, which we term implicit 
electron approach (cf. to section \ref{subsec:IE}), we consider the electron 
implicitly, i.e. we assume that the three ions move on the ground state BO PES. 
In the second approach, which we call explicit electron approach (cf. to section 
\ref{subsec:EE}), we describe the motion of all particles (three ions and one 
electron) simultaneously, so that the electron will be taken into account 
explicitly. In both sections introducing the two apporaches we tutorially
derive the methods and then we summarize the algorithm. 
In both IE and EE approaches state-selective transition probabilities are 
calculated, and the results and further analyses are shown in section \ref{sec:results}.
Finally, the conclusions are given in section \ref{sec:conc}.
%
\section{Model}
\label{sec:model}
Before demonstrating our two approaches, which will be disscussed in the next section, 
we first present the model which we use for the illustration of our implicit and
explicit electron schemes. In order to keep a clear focus on our approaches to describe
electronic nonadiabaticity, we restrict ourselves here to a simple but physically motivated
collinear reactive scattering model. We emphasize that it is straightforward to apply
our implicit and explicit electron approaches also to ab initio Hamiltonians in full dimensionality.

\begin{figure}
\includegraphics[width=0.48\textwidth]{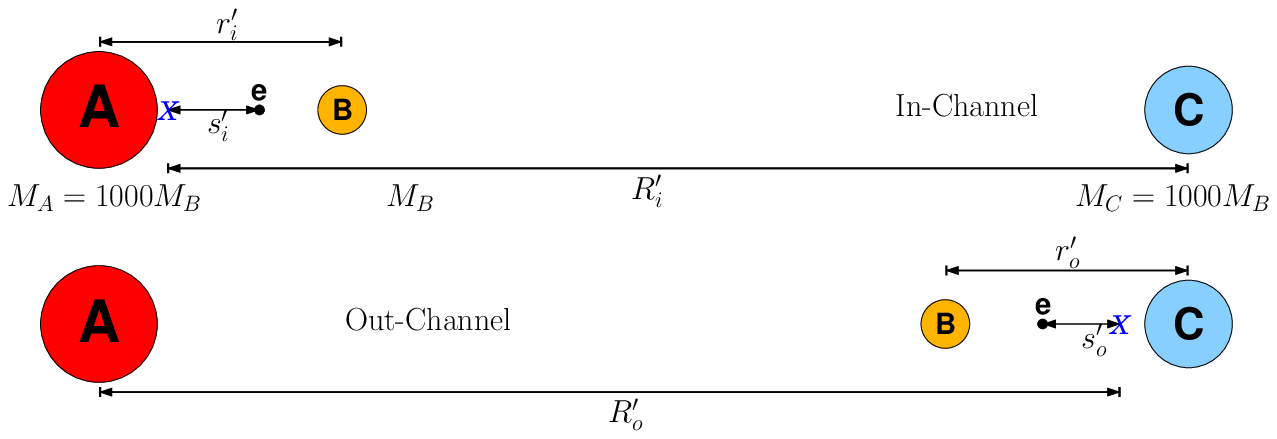}
\caption{\label{fig.model}Schematic representation of our generalization 
of the Shin-Metiu model. In contrast to the original Shin-Metiu model, we allow
all ions to move in the present study. Two asymptotic channels (in and out) are 
considered. The in-channel describes a bound system of ions A, B, and the 
electron. Ion C is initially located far from this bound system. The out-channel 
describes a bound system of ions B, C, and the electron. Here, ion C is 
located far from the bound complex. Transitions from the in-channel to the
out-channel involve a simultaneous electron-ion rearrangement.}
\end{figure}

Our model is quite similar to the original Shin-Metiu model, containing three ions and one 
electron confined to a one-dimensional collinear motion. 
In order to investigate reactive scattering with such a model, we need 
to include the scattering states describing asymptotic channels in which one ion 
is far away from the other part of the system. In other words, we need to remove 
the constraint of fixed terminal ions, to allow all ions to move along the 1D line. 
This generalization of the original Shin-Metiu model is sketched in figure \ref{fig.model}.
Allowing all ions to move enables us to describe the transition from the 
in-channel configuration to the out-channel configuration through a 
collinear collision. This process involves transfers of both an ion and an electron. 

The Hamiltonian of our extended Shin-Metiu model can be written as
\begin{equation}
		\hat{H} = \hat{T}_N  + \hat{T}_e + V_{NN} + V_{eN},
\end{equation}
where $\hat{T}_N$ is the kinetic energy of the three ions
\begin{equation}
		\hat{T}_N =  -\frac{\partial_A^2}{2M_A} - \frac{\partial_B^2}{2M_B} - \frac{\partial_C^2}{2M_C},
\end{equation}
and the kinetic energy of the electron is $\hat{T}_e = -\partial_e^2/2m_e$ 
(atomic units are used throughout). 
Since we allow the ions A and C to move, the masses $M_A$ and $M_C$ are 
in general finite (the infinite mass limit corresponds to the original 
model of fixed ions). Here we focus on a case where $M_A$ and $M_C$ are
large compared to the mass $M_B$ in the center. This allows us to model a
light-atom transfer process. We emphasize that the positions of A and C
could also be viewed as center-of-mass cordinates of small clusters or
nano-particles motivating further a small mass ratio $M_B/M_{A,C}$.
The most important parameter for electronic nonadiabaticity in the
present model is therefore the mass ratio between the middle ion and 
the electron. The masses of the ions at the terminal positions 
do not play an essential role in our discussion as long as they are 
much larger than the mass $M_B$ of the ion in the center. 

In atomic units, we take $M_A=1000M_B$, $M_C=1000M_B$, $m_e=1$. For the
center ion we consider two cases (i) $M_B=1836m_e \sim m_{H}$ and (ii) 
$M_B=3\times1836m_e \sim 3m_{H}$, i.e. the mass of the ion B is taken to be the
proton mass or 3 times the proton mass. It is expected that the electronic 
nonadiabaticity differs in the two cases.
The interaction between the ions is given by $V_{NN}=V_{AB}+V_{BC}+V_{CA}$, 
where we choose short range interactions with the following form
\begin{equation}
		V_{AB} = \frac{h_{AB}\alpha_{AB}^2}{\sinh^2\left(\alpha_{AB} (X_{A}-X_{B})\right)}.
\end{equation}
Here, $h_{AB}$ and $\alpha_{AB}$ are parameters that tune the strength 
and the range of the interaction. We employ similar expressions for 
$V_{BC}$ and $V_{CA}$. With $V_{eN}=V_{eA}+V_{eB}+V_{eC}$ we denote the 
electron-ion interaction which we also choose to be short ranged and 
given by the following form
\begin{equation}
		V_{eA} = -\frac{g_A\beta_A^2}{\cosh^2\left(\beta_A (x_{e}-X_{A})\right)}.
\end{equation}
With $X_{{\{A,B,C\}}}$ and $x_{e}$, we denote the ionic and electronic coordinates
respectively.
Again, $g_A$ and $\beta_A$ are parameters characterizing the strength and 
the range of the potential. Similar expressions are employed for $V_{eB}$ and 
$V_{eC}$.
Such a choice of interaction potentials qualitatively captures a realistic 
situation in which ions are repulsive to each other and are attractive to 
the electron.
The short range potentials chosen here are for the convenience of the scattering 
calculations.

One feature of the chosen potentials that should be highlighted is that the 
ion-ion repulsion is singular at zero separation. This imposes the constraint 
that the ions cannot bypass each other, or in other words, the ions preserve 
the order during the scattering process. Hence, we only need to consider two 
asymptotic channels (cf. figure 1).
On the other hand, the electron-ion attraction is soft at zero separation, which
allows the electron to pass the ions. 
In order to illustrate our approach, we choose in the present work for the range 
parameters
$\alpha_{AB}=\alpha_{BC}=\alpha_{CA}=0.70$, and
$\beta_{A}=\beta_{B}=\beta_{C}=1.70$. The interaction strengths are given by $h_{BC}=1.00$, $h_{AB}=h_{CA}=1.002$, 
and $g_{A}=1.002$, $g_{B}=g_{C}=1.00$. The parameters are chosen to produce 
a physical potential-energy surface for rearrangement scattering. Note, that
it is predominantely the masses and not the interaction parameters that determine
the magnitude of electronic nonadiabaticity.

We emphasize that in general the choice of interaction potentials is not 
imposing any restrictions on our approach. The selected potentials and parameters
are physically motivated, keep the present discussion 
simple, and allow us to focus on the central topic of this work, the description
of nonadiabatic electronic motion in electron-ion rearrangement collisions.

\section{Quantum Reactive scattering treatment}
\label{sec:scattering}
In this section, we introduce two quantum reactive scattering approaches
to calculate the transition probabilities
of the rearrangement collision. The electronic nonadiabaticity
will be visualized through the comparison of the two approaches.
In the first approach, 
the information of the electron is only contained in the ground-state
BO PES. We therefore call it {\em implicit electron (IE) approach}.
In the second approach, the electron and the three ions are considered
all at the same level, i.e. we solve a four-body 
quantum reactive scattering problem. In the following we refer to this as
{\em explicit electron (EE) approach}. 
In the two approaches, we calculate the transition (reaction) probability
and thus obtain the reaction rate by taking an average over Boltzmann factors. 
The differences in the two approaches indicate the electronic nonadiabaticity.

The model described in the last section is used to demonstrate the two approaches.
In the following discussion, we only focus on the figures for the case 
$M_B=m_H$ for illustrations. The plots are qualitatively very 
similar for the case of $M_B=3m_H$.

\subsection{Implicit Electron Approach}
\label{subsec:IE}
\subsubsection{Coordinate System}
In the IE approach on the collinear model,
the three ions are moving on the ground-state PES 
determined by the electronic Hamiltonian, so three degrees of freedom
are needed to describe the system. Since there is no external field, 
the system is translationally invariant. Hence, if we choose Jacobi
coordinates, and separate off the degree of freedom describing the 
center-of-mass motion, then only two internal degrees of freedom are 
left, which can be chosen as
$r_{i}'=X_{B}-X_{A}$ and $R_{i}'=X_{C}-(M_AX_A+M_BX_B)/(X_A+X_B)$
for the in-channel configuration or as $r_{o}'=X_{C}-X_{B}$ and
$R_{o}'=(M_BX_B+M_CX_C)/(M_B+M_C)$ for the out-channel configuration.
Each set of coordinates has its merits in describing a particular
configuration of the system. However, in order to describe the whole 
scattering process using one set of coordinates, we employ in the 
following a mass-weighted
hyperspherical coordinate system\cite{kuppermann1980}. To this end,
we first define mass-weighted coordinates as
\begin{subequations}
\begin{gather}
		r_{i}=\sqrt{\frac{\mu_{AB}}{m}}r_{i}', \quad\quad
		R_{i}=\sqrt{\frac{\mu_{C,AB}}{m}}R_{i}', \\
		r_{o}=\sqrt{\frac{\mu_{BC}}{m}}r_{o}', \quad\quad
		R_{o}=\sqrt{\frac{\mu_{A,BC}}{m}}R_{o}',
\end{gather}
\end{subequations}
in which the $\mu$'s denote different reduced masses. For example, 
$\mu_{AB}$ is the reduced mass of
$A$ and $B$,  $\mu_{C,AB}$ is the reduced mass of $C$ and the center-of-mass 
of $AB$. $m$ is an arbitrary mass, we choose it to be equal to $M_{B}$ in this
paper.
The two new sets of coordinates have the noteworthy property 
\begin{equation}
		r_{i}^2 + R_{i}^2 = r_{o}^2 + R_{o}^2,
\end{equation}
which allows us to introduce a polar coordinate system by defining
\begin{subequations}
\begin{gather}
		\rho = \sqrt{r_{i}^2+R_{i}^2} = \sqrt{r_{o}^2+R_{o}^2}	 \\
		\theta = \arctan(r_{i}/R_{i}) = \theta_m - \arctan(r_{o}/R_{o}).
\end{gather}
\end{subequations}
In terms of these new coordinates, the in-channel and out-channel configurations
can be described on equal footing. It can be shown that the 
angle $\theta \in [0,\theta_m]$ is bounded, with
\begin{equation}
		\theta_{m}=\arctan\sqrt{\frac{M_{B}(M_{A}+M_{B}+M_{C})}{M_{A}M_{C}}}.
\end{equation}

\subsubsection{Hamiltonian}
\begin{figure}
       \centering
           \includegraphics[width=0.48\textwidth]{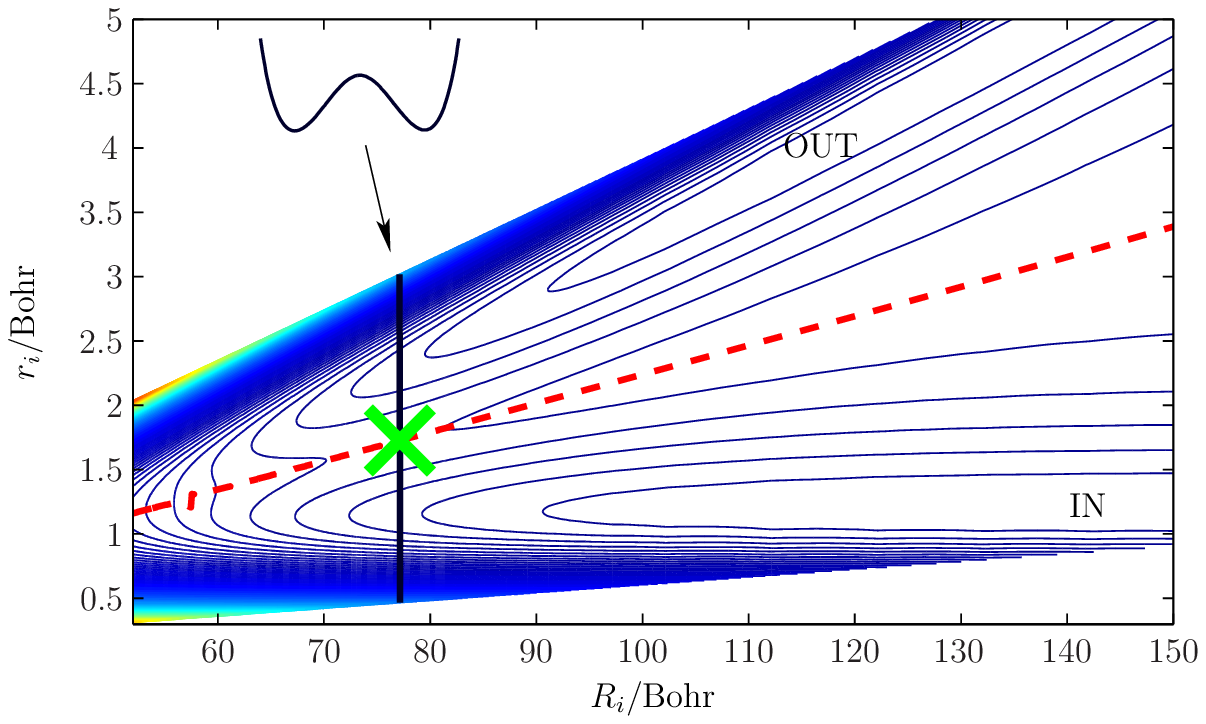}
	   \caption{Ground-state BO surface $V(\rho,\theta)$ for the model ($M_B=m_H$) 
                    in the mass-weighted hyperspherical coordinate system, with the 
                    corresponding Cartesian coordinates $(r_{i},R_{i})$. Note the two 
                    very different scales of the axes. The green cross indicates the 
                    position of the classical transition state. It is the energy 
                    minimum along the red dashed line, and the maximum along the 
                    black solid cut for a constant $\rho$.
	            The shape of $V(\rho,\theta)$ along the black solid cut is 
                    schematically sketched in the inset in the top-left corner. The two 
                    valleys in the plot can be identified to either describe 
                    in-channel or out-channel configurations.}
	   \label{fig.contour}
\end{figure}

The Hamiltonian in the above introduced mass-weighted hyperspherical 
coordinate system can be written as
\begin{equation}
		\hat{H}=-\frac{1}{2m}\left[\frac{\partial^{2}}{\partial\rho^{2}}+
		\frac{1}{\rho}\frac{\partial}{\partial\rho}+\frac{1}{\rho^{2}}\frac{\partial^{2}}{\partial\theta^{2}}\right]+V(\rho,\theta),
\end{equation}
where $V(\rho,\theta)$ is the ground-state BO surface shown in 
figure \ref{fig.contour} for the case $M_B=m_{H}$. 
For every given nuclear configuration,
we solve for the ground-state BO surface by exact diagonalization 
of the electronic Hamiltonian in a finite-difference representation 
in hyperspherical coordinates.

In figure \ref{fig.contour}, it can be seen that for a given radius $\rho$, the 
cut along $\theta$ has a double-well shape, and the minimum corresponds to either 
the in-channel or the out-channel configuration.
From the scale of the coordinate axes it is obvious that the radius $\rho$ is a 
slow variable compared to the angle $\theta$, which implies that the two variables
are approximately decoupled.

We therefore perform the hyperspherical expansion for the eigenstates of the 
Hamiltonian by regarding the radius as a 
parameter\cite{lopez1984,babamov1983_1,babamov1983_2}
\begin{equation}
		\Psi_n(\rho,\theta)=\rho^{-\frac{1}{2}}\sum_{n}\varphi_{n}(\rho)\chi_{n}(\theta;\rho),
		\label{eq.hyper_exp}
\end{equation}
where $\{\chi_{n}(\theta;\rho)\}$ is a set of complete orthonormal functions 
of $\theta$ for a given $\rho$. 
This expansion can be truncated by choosing a proper set of functions
$\left\{\chi_{n}(\theta;\rho)\right\}$.

In order to compute the $\chi_{n}$, we first solve the angular part of the 
Hamiltonian for a given $\rho$
\begin{equation}
		\left[-\frac{1}{2m\rho^{2}}\frac{\partial^{2}}{\partial\theta^{2}}+V(\rho,\theta)\right]\eta_i(\theta;\rho)
		= \epsilon_i(\rho)\eta_i(\theta;\rho), \label{eq.IE_angular}
\end{equation}
where $\eta_{i}(\theta;\rho)$ is the eigenfunction with eigenvalue 
$\epsilon_i(\rho)$ for a given $\rho$.
\begin{figure}
		\centering
		\includegraphics[width=0.48\textwidth]{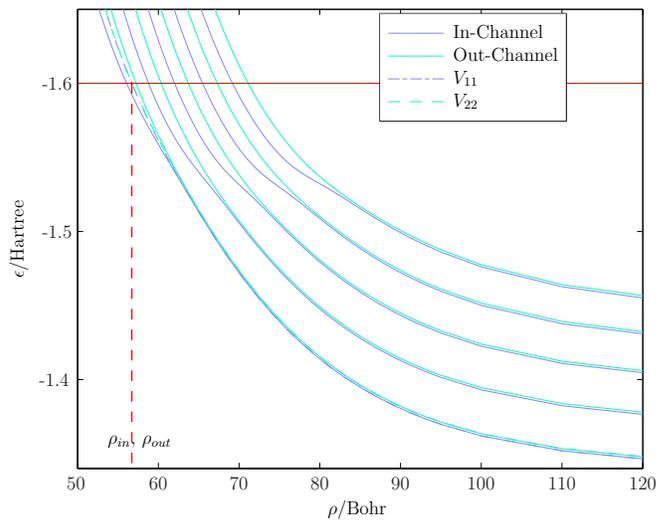}
		\caption{The lowest ten eigenvalues $\epsilon_{i}(\rho)$ of the angular 
                         Hamiltonian (for the case $M_B=m_{H}$) as a function of 
                         the radius are shown with solid lines. The two colors 
                         denote either eigenstates with in-channel (localized in 
                         the in-channel valley) or out-channel-like (localized 
                         in the out-channel valley) character. The dashed-dotted 
                         lines $V_{11}$ and $V_{22}$ are the energies corresponding 
                         to maximally localized states obtained after mixing the 
                         original in-channel and out-channel-like states (see text). }
		\label{fig.a_eigenvalue}
\end{figure}
The eigenvalues $\epsilon_{i}(\rho)$ of the angular Hamiltonian are shown in 
figure \ref{fig.a_eigenvalue}. The curves are plotted in two different colors
depending on whether they are in-channel states or out-channel states, according 
to where the wave function is localized.
The relative position between the two sets of curves is very sensitive to the 
input parameters (masses and interaction strength), and in our case we choose the 
interaction parameters to be asymmetric to avoid degeneracies in these
curves. This allows to unambiguously identify the nondegenerate asymptotic states 
as vibrational states of in- or out-channel configurations.
For our setup, it can be seen that these eigenvalues
appear in pairs.
The wave functions in a given pair also specify the internal vibrational
states of the initial and final scattering wave functions. 
The scattering from a given in-channel configuration to a given out-channel 
configuration can be related to a transtion within a pair. 
Since the two wave functions in a pair span a 2-dimensional space that is 
approximately decoupled from the space spanned by the wave functions belonging 
to other pairs (inter-pair distances are large), by considering each pair 
separately, we can have a state-to-state description of the reaction, 
which is the advantage of reactive-scattering theory. 
To make the illustration as simple as possible, in the following we only take 
the lowest pair, within which the transition gives the main contribution
to the transition probability since this pair is energetically more favourable.

\begin{figure}
		\centering
		\includegraphics[width=0.48\textwidth]{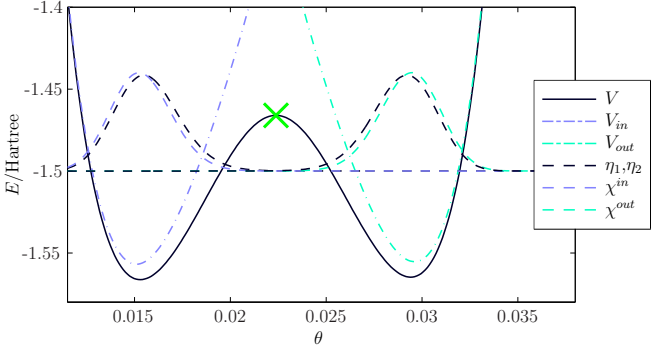}
		\caption{Angular potential $V(\rho,\theta)$ for the case $M_B=m_{H}$ 
                (black solid line) along the cut across the saddle point on the PES.
		The angular wave functions of the lowest pair $\eta_{1}$ and $\eta_{2}$, 
                both in black dashed lines, are localized in 
		one of the two minima of the double well, corresponding to an 
                in-channel (left) or an out-channel (right) state. The maximally 
                localized states $\chi^{in}$ and $\chi^{out}$ obtained by mixing 
                $\eta_{1}$ and $\eta_{2}$ are plotted in dashed lines in different 
                colors. The $V_{in}$ and $V_{out}$ for the localization scheme are 
                plotted with dashed-dotted lines in the same color setting 
                for $\chi^{in}$ and $\chi^{out}$.}
		\label{fig.double_minimum}
\end{figure}
The angular potential $V(\theta;\rho)$ along a fixed $\rho$ cut is shown in 
figure \ref{fig.double_minimum}. The wave functions belonging to the lowest pair
are also given in this figure. The one localized in the left valley ($\eta_1$) 
is an in-channel state, and corresponds to the ground-vibrational state
of the $AB$ molecule (zero nodes). The other one ($\eta_{2}$), localized in the 
right valley, is an out-channel state, and can be interpreted as 
the ground-vibrational state of the $BC$ molecule (zero nodes). 
If we choose $\chi_1$ and $\chi_2$ that span the same space spanned by 
$\eta_1$ and $\eta_2$, then the expansion in Eq. (\ref{eq.hyper_exp})
can be approximately written as a linear combination of $\chi_1$ and $\chi_2$. 
Moreover, if $\chi_1$ and $\chi_2$ are chosen to be maximally localized wave 
functions in one of the valleys, then the full asymptotic wave function 
for in-channel and out-channel configurations can be written as
\begin{gather}
		\Psi^{in}_{\nu=0}(\rho,\theta)=\rho^{-1/2}\varphi_1(\rho)\chi_1(\theta;\rho) \\
		\Psi^{out}_{\nu=0}(\rho,\theta)=\rho^{-1/2}\varphi_2(\rho)\chi_2(\theta;\rho) .
\end{gather}

To get the maximally localized wave functions, which best represent the 
two asymptotic channels, we first define two auxilliary potentials 
$V_{in}$ and $V_{out}$
\begin{subequations}
		\begin{gather}
				V_{in}(\rho,\theta) = V(\rho,\theta) - V_{eC}(\rho,\theta) \\
				V_{out}(\rho,\theta)=V(\rho,\theta)-V_{eA}(\rho,\theta) .
		\end{gather}
\end{subequations}
The potentials $V_{eC}$ and $V_{eA}$ describe the attraction from ion 
C and A respectively. 
The two auxiliary potentials, that have only one valley, are also shown in 
figure \ref{fig.double_minimum}. 
We mix the wave functions $\eta_{1}$ and $\eta_2$ by defining an orthogonal 
transformation
\begin{equation}
		\left(\begin{array}{c}
				\chi_{1}\\
				\chi_{2}
		\end{array}\right)=\mathbf{T}\left(\begin{array}{c}
				\eta_{1}\\
				\eta_{2}
		\end{array}\right)=\left(\begin{array}{cc}
				\cos\zeta & -\sin\zeta\\
				\sin\zeta & \cos\zeta
		\end{array}\right)\left(\begin{array}{c}
				\eta_{1}\\
				\eta_{2}
		\end{array}\right),
\end{equation}
where $\zeta\in[0,\frac{\pi}{2}]$. The parameter $\zeta$ is chosen such that 
the quantity
\begin{equation}
		I=\abs{\braket{\chi^{in}}{\chi_{1}}}^{2}+\abs{\braket{\chi^{out}}{\chi_{2}}}^{2}
\end{equation}
is maximized. The $\chi^{in}$ and $\chi^{out}$ are eigenfunctions of the 
auxiliary potentials, with the same number of nodes as the original 
wave functions $\eta_1$ and $\eta_2$. The physical meaning is that we want 
to mix $\eta_{1}$ and $\eta_{2}$ such that the new wave functions 
maximally overlap with the exact localized functions.

To obtain the transformation matrix $\mathbf{T}$, we regard $\zeta$ as a
variational parameter. Optimizing the localization as function of $\zeta$
yields
\begin{equation}
		I(\zeta)'
		=B\sin2\zeta+2A\cos2\zeta,
\end{equation}
where 
\begin{subequations}
		\begin{gather}
				A=\braket{\chi^{out}}{\eta_{1}}\braket{\chi^{out}}{\eta_{2}}-\braket{\chi^{in}}{\eta_{1}}\braket{\chi^{in}}{\eta_{2}}, \\
				B=\braket{\chi^{in}}{\eta_{2}}^{2}+\braket{\chi^{out}}{\eta_{1}}^{2}-\braket{\chi^{in}}{\eta_{1}}^{2}-\braket{\chi^{out}}{\eta_{2}}^{2}.
		\end{gather}
\end{subequations}
By setting $I(\zeta_{0})'=0$, we get $\tan2\zeta_{0}=-2A/B$. To maximize this quantity,
we have $I(\zeta_{0})''\leq0$, which yields $A\sin2\zeta\geq 0$. 
Since there is arbitrariness in choosing the relative phase of the state 
$\ket{\eta_{1}}$, $\ket{\eta_{2}}$, $\ket{\chi^{in}}$ and $\ket{\chi^{out}}$, we 
fix it by choosing the phase such that $\braket{\chi^{in}}{\chi^{out}}\leq0$ and 
$A\leq0$. Hence, we arrive at
\begin{subequations}
\begin{gather}
		\sin2\zeta=\frac{2A}{\sqrt{4A^{2}+B^{2}}}, \\
		\cos2\zeta=\frac{B}{\sqrt{4A^{2}+B^{2}}},
\end{gather}
\label{eq.zeta}
\end{subequations}
for the optimal parameter $\zeta$.

\subsubsection{Distorted Wave Born Approximation (DWBA)}
We follow the approach described in Ref. \citen{lopez1984,babamov1983_1,babamov1983_2} 
for calculating the transiton probability.
The transition within the lowest pair, i.e. $\nu_{AB}=0 \rightarrow \nu_{BC}=0$, 
is determined by the coupled equations
\begin{subequations}
		\begin{gather}
				\left[-\frac{1}{2m}\frac{\partial}{\partial\rho^{2}}-\frac{1}{8m\rho^{2}}+V_{11}(\rho)-E\right]\varphi_{1}(\rho)=-V_{12}(\rho)\varphi_{2}(\rho), \\
				\left[-\frac{1}{2m}\frac{\partial}{\partial\rho^{2}}-\frac{1}{8m\rho^{2}}+V_{22}(\rho)-E\right]\varphi_{2}(\rho)=-V_{21}(\rho)\varphi_{1}(\rho),
		\end{gather}
		\label{eq.coupled}
\end{subequations}
where
\begin{equation}
		V_{ij}(\rho)=\int d\theta\chi_{i}(\theta,\rho)\left[-\frac{1}{2m\rho^{2}}\frac{\partial^2}{\partial\theta^2}+V(\rho,\theta)\right]\chi_{j}(\rho,\theta).
\end{equation}
By using Eq. (\ref{eq.zeta}), it can be shown that
\begin{subequations}
		\begin{gather}
				V_{12}(\rho)=V_{21}(\rho)=\frac{A}{\sqrt{4A^{2}+B^{2}}}(\epsilon_{1}-\epsilon_{2})>0, \label{eq.coupling} \\
				V_{11}(\rho)=\frac{\epsilon_{1}+\epsilon_{2}}{2}+\frac{B(\epsilon_{1}-\epsilon_{2})}{2\sqrt{4A^{2}+B^{2}}}, \\
				V_{22}(\rho)=\frac{\epsilon_{1}+\epsilon_{2}}{2}-\frac{B(\epsilon_{1}-\epsilon_{2})}{2\sqrt{4A^{2}+B^{2}}},
		\end{gather}
		\label{eq.Vij}
\end{subequations}
where $\epsilon_1$ and $\epsilon_2$ are the angular eigenvalues in the lowest pair 
($\epsilon_1<\epsilon_2$) parametrically depending on the radius $\rho$.
The two diagonal terms $V_{11}$ and $V_{22}$ are plotted in figure 
\ref{fig.a_eigenvalue}.

To arrive at the transition probability, the coupled differential equations could be 
solved e.g.~numerically. However, as we will demonstrate in the following, using
the DWBA allows us to arrive at an {\em analytical} expression for the transition amplitude.
Low-energy scattering events which are the prototypical case for many chemical applications
are covered well in this approximation as known from previous studies\cite{lopez1984,babamov1983_1,babamov1983_2}.
Our analytical result for the amplitude provides therefore a useful and efficient tool 
for the analysis of nonadiabatic effects and the comparision of our IE and EE approaches.

The transition amplitude in DWBA can be written as 
\begin{equation}
		t_{21} = m \int_{0}^{\infty} d\rho \varphi_2^{0}(\rho) V_{12}(\rho) \varphi^{0}_{1}(\rho)
\end{equation}
in which the wave functions $\varphi_{1}^{0}$ and $\varphi_{2}^{0}$ are 
solutions of Eq. (\ref{eq.coupled}) by setting the right-hand sides to zero.
The transition probability can be written as
\begin{equation}
		P_{21}=\sin^{2}(2\pi t_{21}),
		\label{eq.eDWBA}
\end{equation}
known as the exponential DWBA\cite{levine1971}.

To allow for an analytical evaluation of $t_{12}$, some approximations have to
be taken. 
First, we linearize the potential near the classical turning points $\rho_{in}$ 
and $\rho_{out}$ (see figure \ref{fig.a_eigenvalue}), since the major part of 
the contribution to the integral comes from a narrow range near that point
\begin{subequations}
		\begin{gather}
				\left(E- V_{11}(\rho)+\frac{1}{8m\rho^{2}}\right) = \left(\rho - \rho_{in} \right)F_{in} \\
				\left(E- V_{22}(\rho)+\frac{1}{8m\rho^{2}}\right) = \left(\rho - \rho_{out} \right)F_{out}.
		\end{gather}
\end{subequations}
The term $1/8m\rho^2$ is much smaller than $V_{11}$, so the turning points 
$\rho_{in}$ and $\rho_{out}$ are almost the values of $\rho$
at the cross point between the horizontal red line of a given energy and 
the $V_{ii}(\rho)$ of the maximally localized states, 
as shown in figure \ref{fig.a_eigenvalue}. $F_{in}$ and $F_{out}$ are the 
corresponding first-order derivatives with respect to $\rho$ at the turning point. 
We can directly write down the unperturbed wave function, which are given in terms 
of Airy functions, $\mathrm{Ai}(\cdot)$
\begin{subequations}
\begin{gather}
		\varphi_{1}^{0}(\rho)=(2/B_{in})^{1/2}\mathrm{Ai}(-B_{in}(\rho-\rho_{in})),	 \\
		\varphi_{2}^{0}(\rho)=(2/B_{out})^{1/2}\mathrm{Ai}(-B_{out}(\rho-\rho_{out})),
\end{gather}
\end{subequations}
with $B_{j}=(2mF_{j})^{1/3}$. We define the averaged turning point 
\begin{equation}
		\rho_{0} = \frac{\rho_{in}+\rho_{out}}{2}	
\end{equation}
and approximate the coupling $V_{12}$ near this point, as in 
Ref. \citen{lopez1984,babamov1983_1,babamov1983_2}, by
\begin{figure}
		\includegraphics[width=0.45\textwidth]{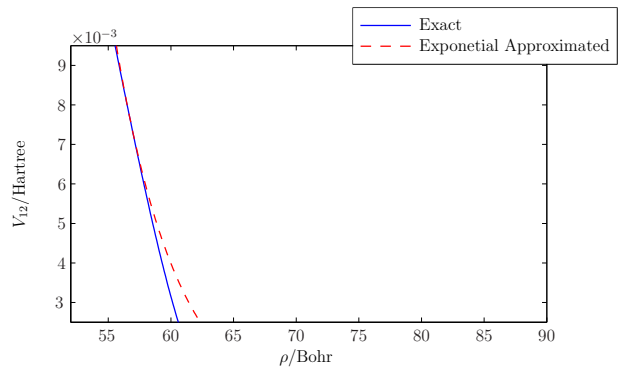}
		\caption{The blue-solid line illustrates the coupling term 
                (for the case $M_B=m_{H}$) obtained by Eq. (\ref{eq.coupling}).
		The red-dashed line shows the exponential approximation for the
		coupling near the averaged turning point $\rho_0$. }
		\label{fig.V12}
\end{figure}
\begin{equation}
		V_{12}(\rho) = V_{12}^{0}e^{-c(\rho-\rho_{0})}
\end{equation}
as shown in figure \ref{fig.V12}. 

Hence, the transition amplitude can be written as
\begin{equation}
		t_{21} \simeq \frac{2mV_{12}^{0}}{\sqrt{B_{in}B_{out}}}\int_{-\infty}^{\infty}dx\mathrm{Ai}(-B_{in}(x-s))\mathrm{Ai}(-B_{out}x)e^{-cx} 
			\label{eq.transition_amplitude}
\end{equation}
where $s=\rho_{in}-\rho_{out}$. We can approximately extend the lower limit 
of the integral to $-\infty$ due to the exponential decay of the two Airy functions. 
The integral can be evaluated analytically, obtaining
\begin{subequations}
\begin{align}
		t_{21} 	=\frac{2mV_{12}^{0}}{\sqrt{B_{in}B_{out}}}\frac{e^{\sigma}{\rm Ai}(\xi)}{\abs{B_{in}^{3}-B_{out}^{3}}^{1/3}},
\label{eq.transition_amplitude_final}
\end{align}
where $\sigma$ and $\xi$ are given by
\begin{gather}
		\sigma =
		\frac{c^{3}(B_{in}^{3}+B_{out}^{3})}{3(B_{in}^{3}-B_{out}^{3})^{2}}-\frac{csB_{in}^{3}}{B_{in}^{3}-B_{out}^{3}} \\
		\xi = \frac{c^{2}B_{in}B_{out}}{(B_{in}^{3}-B_{out}^{3})^{4/3}}-\frac{sB_{in}B_{out}}{(B_{in}^{3}-B_{out}^{3})^{1/3}}.
\end{gather}
		\label{eq.airy}
\end{subequations}
The derivation of this expression is shown in detail in the appendix \ref{app.airy}.
Equation \ref{eq.transition_amplitude_final} is one of the central results in the 
present work. It allows us to express state-to-state resolved transition probabilities
directly in terms of a linearized solution around the averaged classical turning point $\rho_0$.
In the appendix we also demonstrate that in the limit of a symmetric model our result 
in Eq. \ref{eq.transition_amplitude_final} reduces to the well known expression of 
Marcus and coworkers for proton transfer\cite{lopez1984}.

In summary, in practical calculations the IE scheme amounts to the following steps:
\begin{itemize}
	\item Calculate the ground-state PES $V(\rho,\theta)$ and the auxilliary 
              potentials $V_{in}(\rho,\theta)$ and $V_{out}(\rho,\theta)$ 
	      in the hyperspherical coordinates, which can be obtained
	      from any ab-initio method. This is the most time consuming step. 
	\item For each radius $\rho$, solve the angular eigenvalue equation 
              (\ref{eq.IE_angular}) with $V(\rho,\theta)$, and choose the
	      pairs that are of interest for state-selective rates. 
	\item Solve the auxilliary angular eigenvalue equation with 
              $V_{in}(\rho,\theta)$ and $V_{out}(\rho,\theta)$, and use the 
              corresponding states to compute $V_{11}$, $V_{22}$ and $V_{12}$ 
              according to equation (\ref{eq.Vij}).
	\item Use equations ({\ref{eq.eDWBA}}) and (\ref{eq.airy}) to calculate 
              the transition amplitude and probability.
\end{itemize}

\subsection{Explicit Electron Approach}
So far we have discussed an implicit electron approach, where 
the contribution of the electron is only taken into account 
through the BO potential energy surface. In this section we consider
an explicit electron (EE) approach which does not rely on ground-state BO 
surfaces and treats the electron on equal footing with the nuclei during the
scattering process.

\label{subsec:EE}
\subsubsection{Coordinate System}
In the EE approach, after separating off the center-of-mass motion,
we have three degrees of freedom describing the relative motion of the 
system containing three ions and one electron. Similar to the IE
approach, the mass-weighted hyperspherical coordinate system is chosen.
In particular, we first define the mass-weighted Jacobi coordinates for 
the in-channel
\begin{subequations}
\begin{gather}
		r_i =\sqrt{\frac{\mu_{AB}}{m}}\left(X_B - X_A\right)  \\
		s_i = \sqrt{\frac{\mu_{e,AB}}{m}}\left(x_e - \frac{M_A X_A+M_B X_B}{M_A+M_B}\right) \\
		R_i = \sqrt{\frac{\mu_{C,ABe}}{m}}\left(X_C - \frac{M_A X_A+M_B X_B + m_e x_e}{M_A+M_B + m_e}\right) 
\end{gather}
\end{subequations}
and for the out-channel
\begin{subequations}
		\begin{gather}
				r_o = \sqrt{\frac{\mu_{BC}}{m}}\left(X_C - X_B\right) \\
				s_o = \sqrt{\frac{\mu_{e,BC}}{m}}\left(\frac{M_B X_B + M_C X_C}{M_B+M_C} - x_e\right) \\
				R_o = \sqrt{\frac{\mu_{A,BCe}}{m}}\left(\frac{M_B X_B + M_C X_C + m_e x_e}{M_B + M_C + m_e} - X_A\right).
		\end{gather}
\end{subequations}
Since the electron is much lighter than the three ions, both, the $r$ and $R$ 
defined here are almost equal to the ones defined in the IE approach. 

Similar to the IE approach, the two sets of coordinates are related through
\begin{equation}
		r_{i}^{2}+R_{i}^{2}+s_{i}^{2}=r_{o}^{2}+R_{o}^{2}+s_{o}^{2}. 
\end{equation}
Hence, we define the radius as 
\begin{equation}
		\rho = \sqrt{r_{i}^2+R_{i}^2+s_{i}^2}.
\end{equation}
In addition we define two angular arguments. One is similar
to the previous approach,
\begin{subequations}
\begin{gather}
		\theta = \arctan(r_{i}/R_{i}) = \theta_{m} - \arctan(r_{o}/R_{o}) \\
		\theta_{m}=\arctan\sqrt{\frac{m_{2}(M_{1}+m_{2}+M_{3}+1)(m_{2}+M_{3}+1)}{M_{1}M_{3}(m_{2}+M_{3})}}.
\end{gather}
\end{subequations}
The upper bound $\theta_{m}$ is almost the same as the one defined in the 
context of the IE approach. 
The other angular argument 
\begin{equation}
		\phi = \arccos(s_{i}/\rho), \quad\quad \phi \in [0,\pi]
\end{equation}
is new here, and to a large extent behaving like the coordinate of the 
electron. In the following, we will see that the many-particle wave function
along the $\phi$ direction is localized near $\phi=\pi/2$, which corresponds 
physically to the situation that the electron is always localized 
between $AB$ or $BC$. 
We emphasize at this point that the choice of the hyperspherical coordinate system 
is not restricted to the 1D case. Similar to other studies in the literature\cite{Aquilanti1997}
it is straightforward to extend the present discussion to the 3D case. However, the
expressions become then much more involved and the presentation is less transparent.
To demonstrate our approach in a clear way we therefore stay in a 1D setting.

\subsubsection{Hamiltonian}
The Hamiltonian in the hyperspherical coordinate system is written as
\begin{equation}
		\hat{H} =-\frac{1}{2m}\frac{\partial}{\partial\rho^{2}}+ \frac{\hat{L}^{2}}{2m\rho^{2}}+V(\rho,\theta,\phi),
\end{equation}
where $\hat{L}$ is the 3D angular momentum operator, and $V$ includes
the potential energy of electron-ion attraction and ion-ion repulsion. 
If we extend the concept of the PES, then $V$ 
is just a surface in $\rho$, $\theta$ and $\phi$, in which $\rho$ and $\theta$
are almost the same as the ones in the IE approach and can be regarded as 
the ion-like coordinates, while $\phi$ is the electron-like coordinate.
It is this ``PES'' that determines the internal motion of the 
four-particle system and leads to the reactive scattering event. 

Like in the IE approach, in this case the radius $\rho$ can be regarded as a 
slow variable compared to the two angular arguments. Hence we can use the same 
ansatz as we did in the IE approach, i.e. we solve the angular Schr{\"o}dinger 
equation for every given $\rho$
\begin{equation}
		\left[\frac{\hat{L}^{2}}{2m\rho^{2}}+V(\rho,\theta,\phi)\right]\eta_{i}(\theta,\phi;\rho) = \epsilon_{i}(\rho)\eta_{i}(\theta,\phi;\rho).
\end{equation}
Also in this case the eigenvalues appear in pairs. Similar as in the IE approach, we take only the lowest pair.

\begin{figure}
		\includegraphics[width=0.45\textwidth]{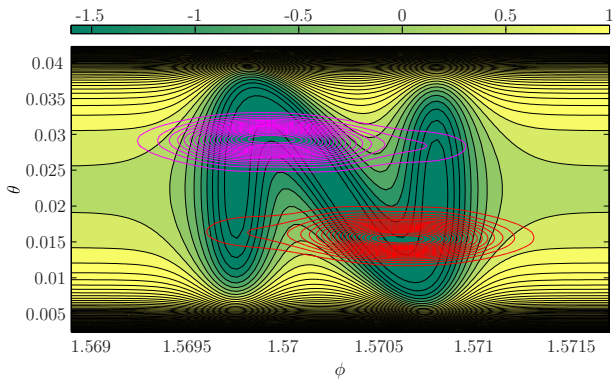}
		\caption{Contour plot of the generalized four-body PES 
                         $V(\rho,\theta,\phi)$ (for the case $M_B=m_{H}$)
		         as a function of $\theta$ and $\phi$ at the same 
                         $\rho$ as in figure \ref{fig.double_minimum}. The 
                         dark region corresponds to the valley in the surface. 
                         The lines in magenta and red are contours of the 
                         wave functions $\eta_{1},\eta_{2}$ respectively, 
                         which correspond to the lowest pair in angular 
                         eigenvalues. }
		\label{fig.full_PES}
\end{figure}

In figure \ref{fig.full_PES}, we show the generalized PES $V(\rho,\theta,\phi)$ as 
a function of the two angular arguments at the same radius as in 
figure \ref{fig.double_minimum}. In addition we show the wave functions from 
the lowest pair. 
Instead of the double-well potential appearing in \ref{fig.double_minimum}, 
here $V(\rho,\theta,\phi)$ contains a long narrow valley with zigzag structure. However, the 
2D angular wave functions from the lowest pair share the same properties 
as the 1D wave functions in the IE approach. In particular,
along $\theta$ direction, the in-channel state is localized at a 
smaller $\theta$ region with zero nodes,
while the out-channel state is localized with zero nodes in the region 
with a larger $\theta$.
Along the other coordinate, the wave functions are sharply localized 
near the region $\phi=\pi/2$. From a geometric perspective, the wave functions 
are strongly confined to the equatorial plane of the sphere described 
by $(\rho,\theta,\phi)$.  
The IE approach is obtained effectively by neglecting the smearing of 
the wave function out of the equatorial plane. 
Since $\phi$ to a large extent is like an electronic coordinate, we can 
regard this spread of the wave function as the origin of 
electronic nonadiabaticity.
In figure \ref{fig.full_PES}, it can be seen that along $\phi$, both 
wave functions have zero nodes. 
This is just because the electron is in the ground state in both channels.  

In the following, we can apply the same localization scheme to the 
wave functions $\eta_{1},\eta_2$ as we did in the IE approach in order 
to obtain the maximally localized states $\chi_1,\chi_2$ and the coupling 
term $V_{12}$. Then, as before, the transition probability can be 
calculated using the DWBA.

\subsubsection{Distorted Wave Born Approximation (DWBA)}
In this section, we calculate the transition probability within the lowest
pair, which is determined by the coupled equations
\begin{subequations}
		\begin{gather}
				\left[-\frac{1}{2m}\frac{\partial}{\partial\rho^{2}}+V_{11}(\rho)-E\right]\varphi_{1}(\rho)=-V_{12}(\rho)\varphi_{2}(\rho) \\
				\left[-\frac{1}{2m}\frac{\partial}{\partial\rho^{2}}+V_{22}(\rho)-E\right]\varphi_{2}(\rho)=-V_{21}(\rho)\varphi_{1}(\rho),
		\end{gather}
\end{subequations}
where  $V_{ij}(\rho)$ is given by Eq.~(\ref{eq.Vij}), since we have taken 
exactly the same localization scheme as we did in the IE approach. 

To calculate the transition amplitude $t_{21}$ analytically, the same 
approximations for $V_{ij}$ are taken. Thus, without any difficulties,
we obtain the transition probability $P_{21}$ by using Eq.~(\ref{eq.eDWBA}).

Compared to the IE approach, for the EE approach we need to 
perform the following steps in practice:
\begin{itemize}
	\item Calculate the generalized PES $V(\rho,\theta,\phi)$, 
              the auxilliary potentials $V_{in}(\rho,\theta,\phi)$
	      and $V_{out}(\rho,\theta,\phi)$ for all particles (ions 
              and the electron) at different configurations in the
	      hyperspherical coordinates. This is the most time consuming step.
	\item For different radii $\rho$, solve the angular eigenvalue 
              equation with the generalized PES $V$. Then choose the pairs 
              that are of physical interest.
	\item Solve the auxilliary angular eigenvalue equation with 
              $V_{in}$ and $V_{out}$, and use the corresponding states
	      to compute $V_{11}$, $V_{22}$ and $V_{12}$ according to 
              equation (\ref{eq.Vij}). This is the same as in the IE approach.
	\item Use the same formulae, i.e.~Eq. ({\ref{eq.airy}}) and 
              (\ref{eq.eDWBA}) to calculate the transition amplitude and probability.
\end{itemize}

\section{Results}
\label{sec:results}
\begin{figure}
	\centering
	\includegraphics[width=0.48\textwidth]{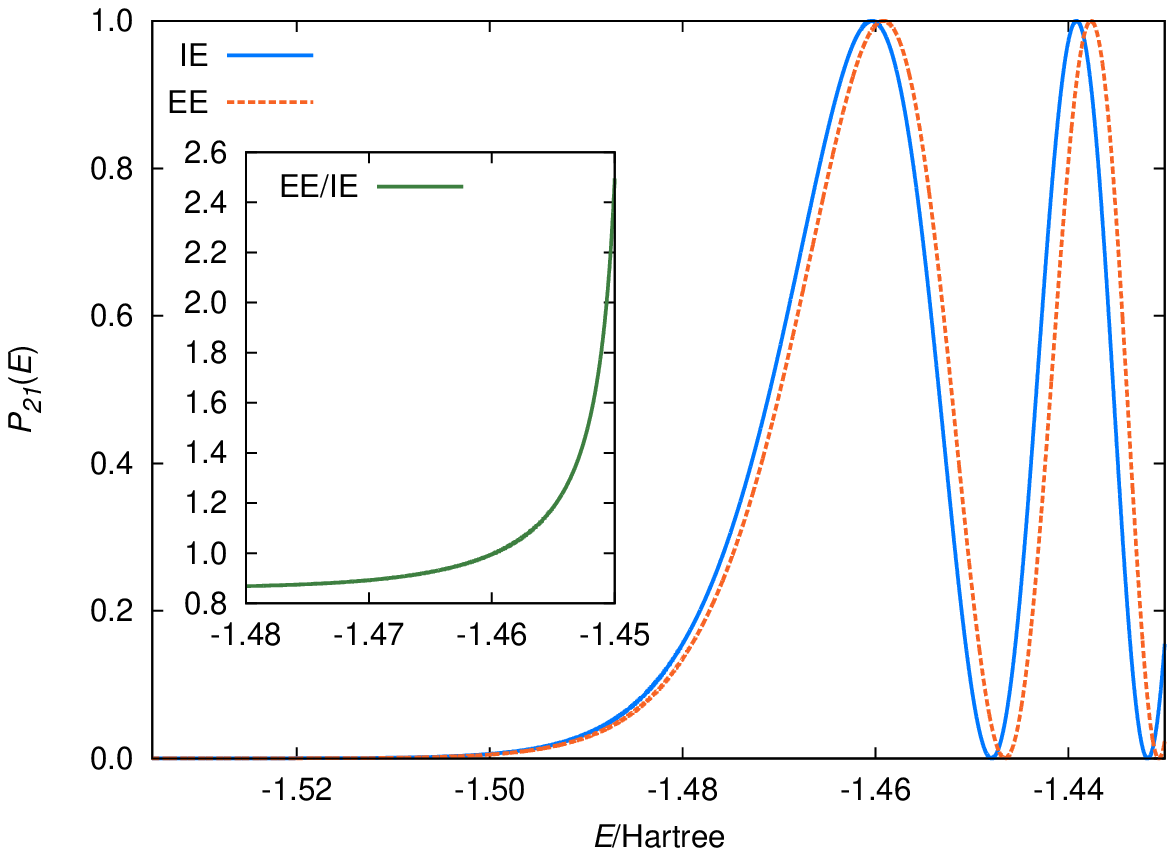} 
	\includegraphics[width=0.48\textwidth]{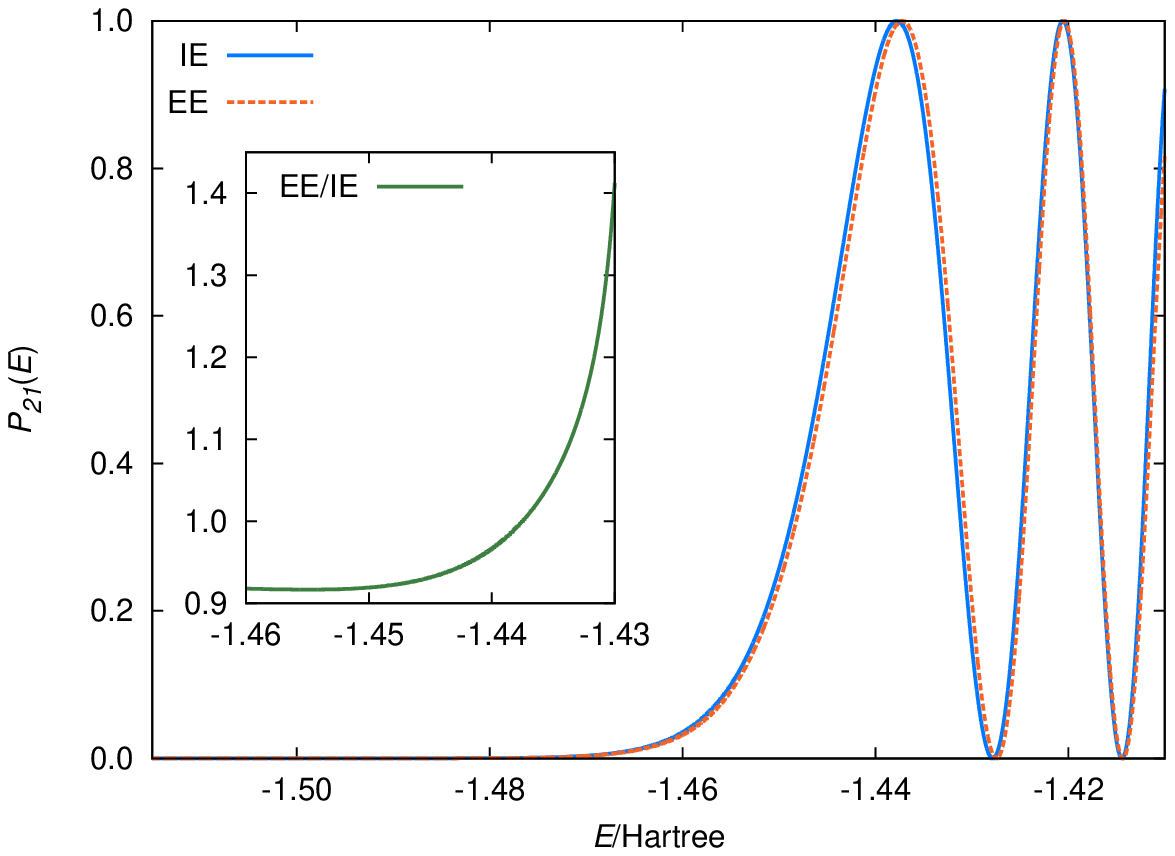} 
		\caption{\label{fig.Probability}Transition probability 
                         $\nu_{AB}=0$ $\rightarrow$ $\nu_{BC}=0$
			 as a function of the energy of the incident ion C as 
			 obtained from the approaches of IE (blue) and EE (red). 
			 The inset in each panel is the ratio of the probabilities, 
                         obtained from the approaches of EE and IE.
			 The upper panel shows the case $M_B=m_H$, and the 
			 lower panel displays $M_B=3m_H$.}
\end{figure}

To visualize the electronic nonadiabaticity in the reactive scattering context, 
we compare the different probabilities obtained from the IE and EE approaches 
described above, for the two cases $M_B=m_H$ and $M_B=3m_H$, 
which are shown in figure \ref{fig.Probability}. The horizontal axis is the total 
energy of the system, which can be tuned by changing the incident kinetic energy 
of ion C. Since we treat the IE and EE approaches completely in parallel, the 
main difference between the schemes arises from whether the electron is treated explicitly or not. 
There are small deviations from the exact case due to the truncation in the hyperspherical expansion
and the two-state approximation.
However, these approximations become exact as the mass ratio between the ions in the 
middle and at the ends approaches zero. For the mass ratios $M_B/M_{A,C}$ 
considered here, which are at the order of $10^{-3}$, the approximations are very accurate.
Thus the deviation in the transition probability between the two approaches can 
almost exclusively be attributed to the difference between IE and EE, which is 
the main contribution of the electronic nonadiabaticity.

In figure \ref{fig.Probability}, the probability is plotted in a relatively 
low-energy regime, because the linear and exponential approximation for 
$V_{ij}(\rho)$ is only valid for low-energy scattering 
calculations\cite{lopez1984,babamov1983_1,babamov1983_2}. From the figure, it 
is obvious that the rearrangement happens when the energy is above some threshold. 
In other words, only when the ion C has enough kinetic energy, the collision 
leads to a reactive rearrangement.

In low energy scattering regime, which is of chemical interests,
it is shown that as the energy increases, the difference between the results from the two 
approaches becomes larger and larger. This directly shows the electronic
nonadiabaticity is pronounced when the kinetic energy of the ions is 
relatively large. 
By comparing the results for $M_B=m_H$ and $M_B=3m_H$, we 
see that the electronic nonadiabaticity is more pronounced when the mass of the 
central ion, or the mass ratio between the central ion and the electron, 
is smaller. This directly reflects the fact that the nonadiabaticity comes 
from the coupled motion of the electron and central ion during the 
rearrangement. The larger the mass ratio is, the better the traditional 
BO description, or similarly, the IE approach will be, as expected.

\begin{figure}
\includegraphics[width=0.48\textwidth]{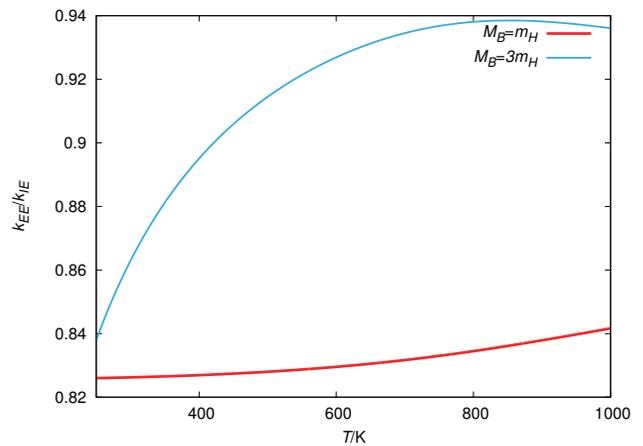}
\caption{\label{fig.rate_ratio}
  The ratio $k_{EE}/k_{IE}$  of the reaction rates for $M_B=m_H$ and $M_B=3m_H$ 
  from the EE and IE approaches as a function of temperature $T$.}
\end{figure}

We also notice that the probability obtained in the IE approach increases 
faster from zero than the one in the EE approach. This implies that the rate 
of the rearrangement is slower in the EE approach. To see this, we calculate 
the ratio of the reaction rates from the two approaches for a certain range of 
temperatures, which is shown in figure \ref{fig.rate_ratio}. We have taken 
the standard expression\cite{Henriksen2008} that assumes the rate is 
the canonical average of many collisions 
where the kinetic energy of the incident particle C is taken from a 
canonical distribution at temperature $T$: 
\begin{equation}
		k(T)\simeq \frac{1}{2\pi Z_{in}}\int_{0}^{+\infty}P_{21}(E+E_0)\exp{(-\beta E)}dE,
\end{equation}
where $Z_{in}$ is the vibrational partition function of the in-channel configuration
and $E_0$ denotes the ground-state energy of the in-channel wavefunction. Here we 
approximate the contribution from transitions between all pairs by only the lowest pair.
This is valid because from figure \ref{fig.a_eigenvalue}, the gap between the lowest pair and the 
pair from the 1st excited states in the asymptotic region is approximately 0.3 Hartree, 
which means in thermodynamic equilibrium, the population ratio between the the second lowest pair
and the lowest pair according is about $\exp(-100)$, which is negligible. 
An alternative assumption is that the kinetic energy of the incident particle
C is sharply peaked at a particular value of energy. The ratio of the reaction 
rates corresponds to the ratio of the two $P_{21}(E)$ curves. 
This ratio is shown in the insets of both panels of Fig.
\ref{fig.Probability} for values of $E$ where $P_{21}(E)$ is non negligible.

The behaviour of the ratio of the reaction rates can be understood thinking that 
the EE approach goes beyond the usual ground-state BO approximation and effectively considers the electronic 
excited states. Since the first excited-state surface often has a positive 
curvature near the maximum of the ground-state surface, as in this model\cite{shin1995}, 
this implies that the excited state is energetically repulsive along the direction 
leading to the rearrangement. This means that any wavepacket with population restricted
to the first excited state will always be bounced back, leading 
to nonreactive scattering.
The result in the IE approach, or ground-state BO approximation, overestimates 
the rate which coincides with the results in the original paper of the 
Shin-Metiu model, which relies on a quite different approach\cite{shin1995}. Further, 
from comparing the results for the two cases with different $M_B$, again 
we see that the difference in the rate becomes larger when the central 
ion is lighter.

\begin{figure}
	\centering
	\includegraphics[width=0.45\textwidth]{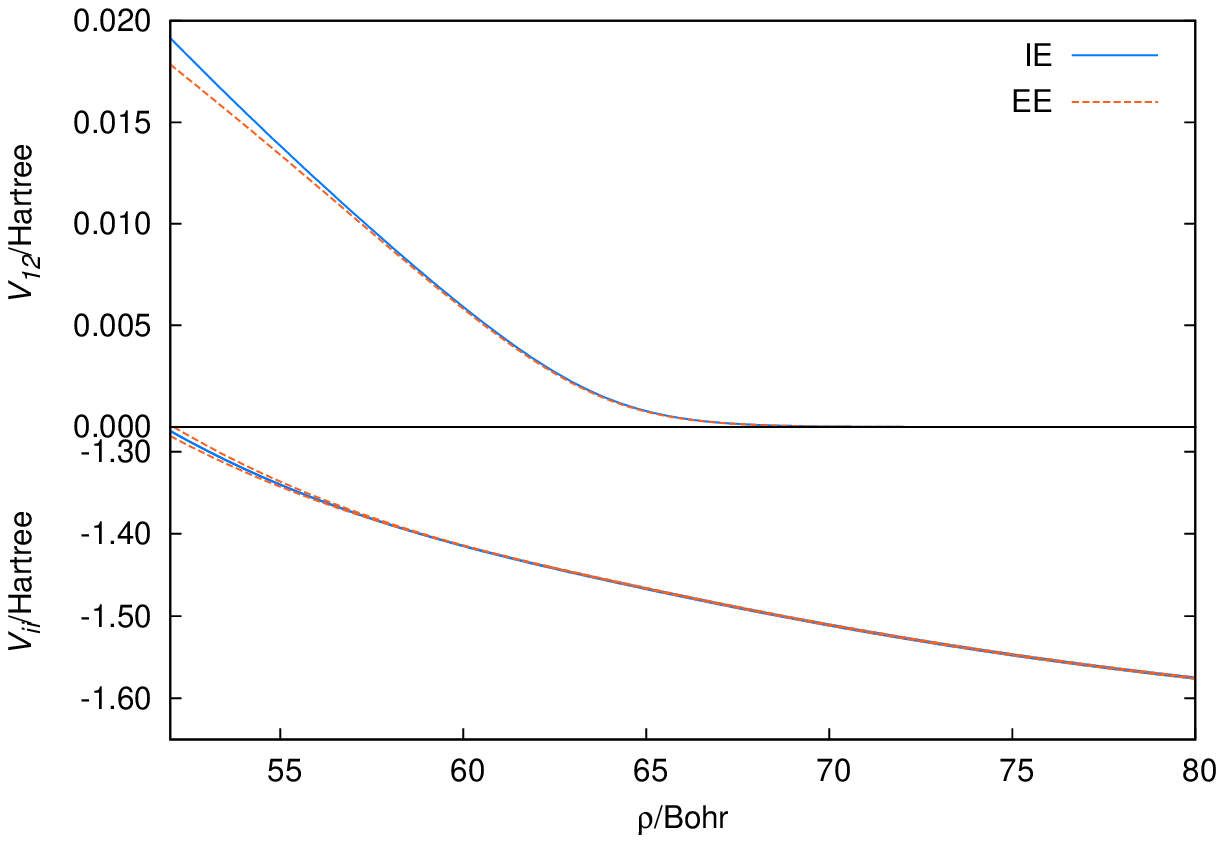}
	\includegraphics[width=0.45\textwidth]{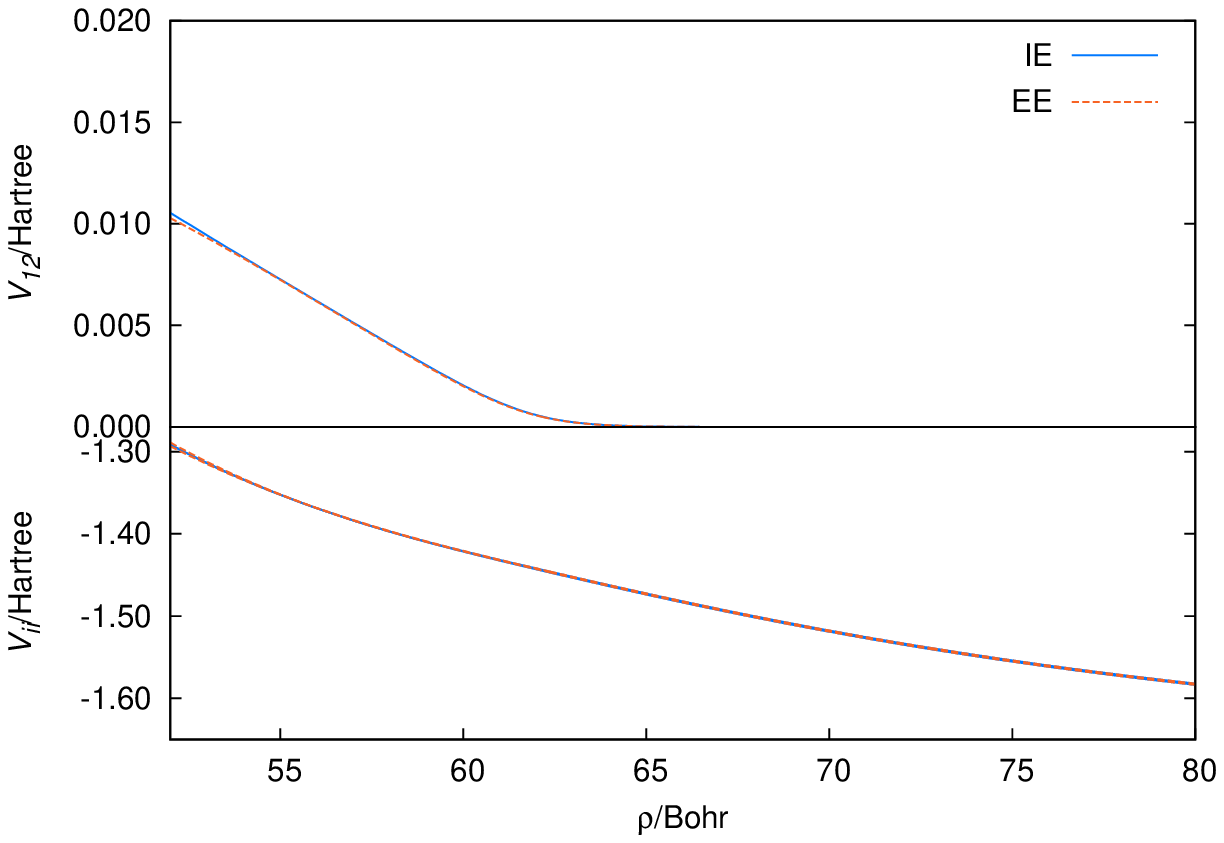}
	\caption{\label{fig.V_compare} 
             The potentials $V_{12}(\rho)$, $V_{11}(\rho)$, and $V_{22}(\rho)$ 
             as a function of radius $\rho$ obtained from the approaches of 
             IE (blue solid) and EE (red dashed). The upper panel displays the 
             case for $M_B=m_H$, and the lower one for $M_B=3m_H$.}
\end{figure}

In both the IE and EE approaches, the potentials $V_{11}(\rho), V_{22}(\rho)$,
and $V_{12}(\rho)$ are used as input for the DWBA calculation, which are crucial 
in determining the transition probability. Hence, in order explore the region 
at which the electronic nonadiabaticity is important, we compare the potentials 
obtained from the two approaches in figure \ref{fig.V_compare} for the two 
cases with different $M_B$.
From these figures, we notice that at large $\rho$, which is the situation when 
one ion is far apart from the other two ions and the electron, the curves from 
the two approaches are on top of each other. Whereas in the region of smaller $\rho$, 
the three ions get together relatively closely, or in other words near the 
classical transition state, there are differences between the two. For 
the $M_B=m_H$ case, the difference is more pronounced. This shows the electronic 
nonadiabaticity is most important near the barrier region or the transition state 
of a chemical reaction, while it hardly contributes to the property of the system 
in the equilibrium geometry, i.e.~in the two asymptotic channels. It also shows 
the nonadiabaticity depends on the mass ratio of the central ion and the electron 
significantly. Hence, this confirms that our EE approach allows to capture electronic 
nonadiabaticity and the resulting renormalization of reaction barriers.

\section{Concluding Remarks}
\label{sec:conc}
In this work we proposed a scheme to capture electronic nonadiabaticity from 
a reactive scattering perspective. 
For reactive rearrangement collisions, we introduced two approaches, one which 
treats the electron implicitly, and a nonadiabatic approach where the 
electron is treated explicitly and on a similar footing as the involved ions. Both
approaches rely on a mass-weighted hyperspherical coordinate system which
allows for an efficient and unified representation of in- and out-channels. In particular,
for the explicit electron approach the transformation to the hyperspherical 
coordinate system introduces a mixing of the original Cartesian coordinates of 
the ions and the electron, which allows to introduce approximations
with smaller error compared to the original Cartesian coordinate system.
Both approaches differ only in the way the electron is treated. We can therefore
conclude that the differences in reaction rates obtained from both methods 
directly reflect the electronic nonadiabaticity. 

To exemplify our approach, the original 
Shin-Metiu model was generalized by removing the constraint of fixed terminal ions.
Within this model, we have investigated two cases, in which the mass of the central 
ion were set to the proton mass and three times the proton mass, while all the other 
system parameters were kept identical. It was shown that the electronic nonadiabaticity 
is larger when the mass ratio between the central ion and the electron becomes 
smaller.
In the frame of the hyperspherical coordinate system, we found that 
nonadiabatic effects are much more pronounced at a small radius $\rho$.
Physically this corresponds to regions near the reaction barrier or transition 
state. In contrast, nonadiabatic effects play no essential role near the equilibrium 
configuration in the asymptotic channels. This illustrates that our proposed approaches 
provide an accurate description for low-energy scattering 
events, which is the typical case for chemical applications. In our investigation, 
we showed that the electronic nonadibaticity leads to a larger deviation in transition 
probabilities as the energy of the system is increasing. We also showed that 
the implicit electron approach overestimates the reaction rate at room 
temperature, since the transition probability increases faster from zero 
compared to the transition probabilities from the explicit electron approach. 
This observation is in accord with the results in earlier work\cite{shin1995}.

The case of extremely high kinetic impact energies, which is of less interest for
common chemical reactions, has not been discussed in the present work. However, we note 
that the nonadiabaticity in this case can be expected to play a minor role, 
since the scattering is fully kinetic. Collisions in this limit are fully 
elastic and thus do not depend at all on the intrinsic electronic structure. 

The present study differs from other studies on nonadiabaticity in that the 
electronic nonadiabaticity was studied from a reactive-scattering perspective, 
and the way we treat the ions and the electron at the same footing is 
conceptually different from other nonadiabatic treatments based on multi-PESs. 
Since the quantum reactive-scattering approach describes reactions at a 
state-to-state resolution (in this paper we consider $\nu_{AB}=0 
\rightarrow \nu_{BC}=0$), our approach allows to gain some insights and
understanding of electronic nonadiabatic effects at a more microscopic 
state-to-state level.

Our way of studying the electronic nonadiabaticity can be utilized to investigate 
real chemical systems involving a light ion which transfers in concert with one 
electron in a collinear arrangement, since the approach does not depend on the 
mathematical form of the interactions. Future prospects include the generalization 
to multi-electron transfers which appear in negative-$U$ systems, where electron 
transfers occur pairwise, or an extension of the present work to more sophisticated
proton-coupled electron transfer reactions as studied in Refs. \cite{Fernandez2013, Auer2012, Fernandez2012}. 
Also an embedding of the present scheme into 
density-functional approaches is desirable. Here further studies are required 
to analyze how such effective (multi-component) density-functional potentials 
have to be constructed.

All these aspects will be in the focus of future studies.

\acknowledgments
This research was supported by the international Max Planck research school for 
complex surfaces in material science (IMPRS-CS). The authors thank Professor 
Matthias Scheffler for his support and useful discussions and Professor John Tully 
for useful discussions during the preparation of the manuscript.\\

\appendix*
\section{Derivation of Eq. (\ref{eq.airy})}
\label{app.airy}
To calculate the transition amplitude, we need to evaluate an integral 
of the following form:
\begin{align}
		I&=\int_{-\infty}^{+\infty}dx{\rm Ai}(-b_{1}(x-c_{1})){\rm Ai}(-b_{2}(x-c_{2}))e^{-ax} \nonumber \\
		&=\left(\frac{1}{2\pi}\right)^{2}\int_{-\infty}^{+\infty}dxdydz\exp\left[if(x,y,z)\right]
\end{align}
in which
\begin{equation}
		f(x,y,z) =\frac{1}{3}(y^{3}+z^{3})+b_{1}c_{1}y+b_{2}c_{2}z-(b_{1}y+b_{2}z)x+iax.
\end{equation}
By introducing new variables
\begin{subequations}
\begin{gather}
			u = y+z\\
			v = b_{1}y+b_{2}z, 
\end{gather}
\end{subequations}
the original variables $y$ and $z$ can be written in terms of the new variables 
\begin{subequations}
		 \begin{gather}
				 y  =\frac{v-b_{2}u}{b_{1}-b_{2}}\\
				z  =\frac{b_{1}u-v}{b_{1}-b_{2}}.
		\end{gather}
\end{subequations}
Thus, we have
\begin{align}
		\frac{1}{3}(y^{3}+z^{3}) &=\frac{u^{3}}{3}-u\frac{(b_{1}+b_{2})uv-v^{2}-b_{1}b_{2}u^{2}}{(b_{1}-b_{2})^{2}} \nonumber \\
		&=\frac{b_{1}^{2}+b_{2}^{2}+b_{1}b_{2}}{3(b_{1}-b_{2})^{2}}u^{3}+\frac{u}{(b_{1}-b_{2})^{2}}\left[v^{2}-(b_{1}+b_{2})uv\right]
\end{align}
\begin{equation}
		b_{1}c_{1}y+b_{2}c_{2}z=\frac{(b_{1}c_{1}-b_{2}c_{2})}{b_{1}-b_{2}}v-\frac{b_{1}b_{2}(c_{1}-c_{2})}{b_{1}-b_{2}}u
\end{equation}
and the Jacobian takes the form
\begin{equation}
		\abs{\frac{\partial(u,v)}{\partial(y,z)}}=\abs{\det\left(\begin{array}{cc}
		1 & 1\\
		b_{1} & b_{2}
\end{array}\right)}=\abs{b_{2}-b_{1}}.
\end{equation}
Hence, The original integral can be expressed as
\begin{equation}
		I=\left(\frac{1}{2\pi}\right)^{2}\int_{-\infty}^{+\infty}\frac{dxdudv}{\abs{b_{1}-b_{2}}}\exp(ig(x,u,v)) \label{eq.int}
\end{equation}
with
\begin{widetext}
\begin{align}
		g(x,u,v)  &= \frac{b_{1}^{2}+b_{2}^{2}+b_{1}b_{2}}{3(b_{1}-b_{2})^{2}}u^{3}-\frac{b_{1}b_{2}(c_{1}-c_{2})}{b_{1}-b_{2}}u+iax \nonumber \\
		&+ c{u}{(b_{1}-b_{2})^{2}}\left[v^{2}-\left((b_{1}+b_{2})u+\frac{((b_{1}-b_{2})x-(b_{1}c_{1}-b_{2}c_{2}))(b_{1}-b_{2})}{u}\right)v\right]
\end{align}
\end{widetext}
Let
\begin{equation}
t=((b_{1}-b_{2})x-(b_{1}c_{1}-b_{2}c_{2}))(b_{1}-b_{2}) ,
\end{equation}
then
\begin{subequations}
		\begin{gather}
				x=\frac{t}{(b_{1}-b_{2})^{2}}+\frac{b_{1}c_{1}-b_{2}c_{2}}{b_{1}-b_{2}} \\
				dx=\frac{dt}{(b_{1}-b_{2})^{2}} .
		\end{gather}
\end{subequations}
By applying the Gaussian integration formula
\begin{equation}
		\int_{-\infty}^{+\infty}dv\exp(i\alpha(v^{2}-\beta v))=\sqrt{\frac{i\pi}{\alpha}}\exp\left(-\frac{i\alpha\beta^{2}}{4}\right)
\end{equation}
to Eq. (\ref{eq.int}), we first integrate over $v$ and find
\begin{equation}
		I=\left(\frac{1}{2\pi}\right)^{2}\int_{-\infty}^{+\infty}\frac{dtdu}{\abs{b_{1}-b_{2}}^{3}}\sqrt{\frac{i\pi}{\alpha}}\exp(ih(t,u))
\end{equation}
with
\begin{equation}
				\alpha=\frac{u}{(b_{1}-b_{2})^{2}} .
\end{equation}
and the function $h(t,u)$ is
\begin{align}
		h(t,u) &= \left[\frac{b_{1}^{2}+b_{2}^{2}+b_{1}b_{2}}{3(b_{1}-b_{2})^{2}}-\frac{(b_{1}+b_{2})^{2}}{4(b_{1}-b_{2})^{2}}\right]u^{3}
		 \nonumber \\
		&-\frac{1}{4u(b_{1}-b_{2})^{2}}\left[t^{2}+\left(2u^{2}(b_{1}+b_{2})-4iau\right)t\right] \nonumber \\
		&-\frac{b_{1}b_{2}(c_{1}-c_{2})}{b_{1}-b_{2}}u+\frac{ia(b_{1}c_{1}-b_{2}c_{2})}{b_{1}-b_{2}}.
\end{align}
Next, we integrate over $t$, and obtain
\begin{align}
		I&=\left(\frac{1}{2\pi}\right)^{2}\exp\left[-\frac{a(b_{1}c_{1}-b_{2}c_{2})}{b_{1}-b_{2}}\right]\nonumber \\
		&\times \int_{-\infty}^{+\infty}\frac{du}{\abs{b_{1}-b_{2}}^{3}}\sqrt{\frac{i\pi}{\alpha}}\sqrt{\frac{\pi}{i\gamma}}\exp(iq(u))
\end{align}
with
\begin{equation}
		\gamma=\frac{1}{4u(b_{1}-b_{2})^{2}}
\end{equation}
and the function $q(u)$
\begin{align}
		q(u) &= \left[\frac{b_{1}^{2}+b_{2}^{2}+b_{1}b_{2}}{3(b_{1}-b_{2})^{2}}\right]u^{3}-\frac{ia(b_{1}+b_{2})}{(b_{1}-b_{2})^{2}}u^{2} \nonumber \\
		&-\left[\frac{b_{1}b_{2}(c_{1}-c_{2})}{b_{1}-b_{2}}+\frac{a^{2}}{(b_{1}-b_{2})^{2}}\right]u.
		\label{eq.q}
\end{align}
By inserting the expressions for $\alpha$ and $\gamma$, the integral $I$ can be simplified to 
\begin{align}
		I &= \left(\frac{1}{2\pi}\right)\exp\left[-\frac{a(b_{1}c_{1}-b_{2}c_{2})}{b_{1}-b_{2}}\right]\frac{1}{\abs{b_{1}-b_{2}}} \nonumber \\
		&\times \int_{-\infty}^{+\infty}du\exp(iq(u)).
\end{align}
At this point we still need to perform the integral over $u$.
To this end, let us first consider another integral of the following form
\begin{equation}
		\tilde{I}=\frac{1}{2\pi}\int_{-\infty}^{+\infty}dx\exp\left[i(Ax^{3}-Bx^{2}-Cx)\right] 
		\label{eq.tildeI}
\end{equation}
with $A>0$. We introduce the variable $x=w+s$ in order to change the integration variable, then 
\begin{align}
		\tilde{I}&=\frac{1}{2\pi}\exp[is(As^{2}-Bs-C)]  \nonumber \\
		&\int_{-\infty}^{+\infty}dw\exp\left[i(Aw^{3}+(3As-B)w^{2}+(3As^{2}-2Bs-C)w)\right].
\end{align}
To eliminate the quadratic term, we set $s=B/3A$. Thus,
this integral can be rewritten as an Airy function
\begin{align}
		\tilde{I} &= \frac{1}{(3A)^{1/3}}\exp\left[-\frac{iB}{3A}\left(\frac{2B^{2}}{9A}+C\right)\right] \nonumber \\
		&\times {\rm Ai}\left(-\frac{1}{(3A)^{1/3}}(\frac{B^{2}}{3A}+C)\right). \label{eq.tildeI_airy}
\end{align}
Now if we look at the expression of the function $q(u)$ in Eq. (\ref{eq.q}), we find that to evaluate the original integral $I$ ,
we just need to calculate the integral $\tilde{I}$ in \ref{eq.tildeI} by taking the parameters
\begin{subequations}
		\begin{gather}
				3A=\frac{b_{1}^{2}+b_{2}^{2}+b_{1}b_{2}}{(b_{1}-b_{2})^{2}}=\frac{b_{1}^{3}-b_{2}^{3}}{(b_{1}-b_{2})^{3}} \\
				B=\frac{ia(b_{1}+b_{2})}{(b_{1}-b_{2})^{2}} \\
				C=\frac{b_{1}b_{2}(c_{1}-c_{2})}{b_{1}-b_{2}}+\frac{a^{2}}{(b_{1}-b_{2})^{2}} .
		\end{gather}
\end{subequations}
Using the result from Eq. (\ref{eq.tildeI_airy}), 
we have
\begin{align}
		\tilde{I}&=\frac{b_{1}-b_{2}}{(b_{1}^{3}-b_{2}^{3})^{1/3}} \nonumber \\
		&\times \exp\left\{ \frac{a^{3}(b_{1}^{3}+b_{2}^{3})}{3(b_{1}^{3}-b_{2}^{3})^{2}}+\frac{ab_{1}b_{2}(b_{1}+b_{2})(c_{1}-c_{2})}{b_{1}^{3}-b_{2}^{3}}\right\}  \nonumber \\
		&\times{\rm Ai}\left(\frac{a^{2}b_{1}b_{2}}{(b_{1}^{3}-b_{2}^{3})^{4/3}}-\frac{b_{1}b_{2}(c_{1}-c_{2})}{(b_{1}^{3}-b_{2}^{3})^{1/3}}\right).
\end{align}
Finally, we arrive at the result for the original integral
\begin{subequations}
\begin{gather}
		I=\frac{e^{\sigma}{\rm Ai}(\xi)}{(b_{1}^{3}-b_{2}^{3})^{1/3}}\quad\quad\quad(b_{1}>b_{2}) \\
		\sigma=\frac{a^{3}(b_{1}^{3}+b_{2}^{3})}{3(b_{1}^{3}-b_{2}^{3})^{2}}-\frac{a(b_{1}^{3}c_{1}-b_{2}^{3}c_{2})}{b_{1}^{3}-b_{2}^{3}}	\\
		\xi=\frac{a^{2}b_{1}b_{2}}{(b_{1}^{3}-b_{2}^{3})^{4/3}}-\frac{b_{1}b_{2}(c_{1}-c_{2})}{(b_{1}^{3}-b_{2}^{3})^{1/3}} .
\end{gather}
\end{subequations}
Although we arrived at an analytical expression for this integral, there is still 
one problem in getting numerical values from this expression when $b_{1}\simeq b_{2}$, 
since the denominator goes to zero and the exponential diverges. Hence, we need to get an asymptotic expression for $b_{1}\rightarrow b_{2}$.
Let
\begin{subequations}
		\begin{gather}
		b_{1}^{3}=b^{3}+\epsilon\\
		b_{2}^{3}=b^{3}-\epsilon
		\end{gather}
\end{subequations}
where $\epsilon \rightarrow 0^+$. Then we have
\begin{gather}
		\sigma=\frac{a^{3}(b_{1}^{3}+b_{2}^{3})}{12\epsilon^{2}}-\frac{a(b_{1}^{3}c_{1}-b_{2}^{3}c_{2})}{2\epsilon} \\
		\xi =\frac{a^{2}b_{1}b_{2}}{(2\epsilon)^{4/3}}\left(1-\frac{2(c_{1}-c_{2})\epsilon}{a^{2}}\right).
\end{gather}
Since as $\epsilon \rightarrow 0$, $\xi \rightarrow +\infty$. We can take the asymptotic expression for the Airy function
\begin{equation}
		{\rm Ai}(\xi)\sim\frac{\exp\left[-\frac{2}{3}\xi^{\frac{3}{2}}\right]}{2\sqrt{\pi}\xi^{1/4}} 
\end{equation}
for $\xi \rightarrow +\infty$. We obtain
\begin{equation}
		\lim_{\epsilon\rightarrow0^{+}}I(\epsilon)=\frac{1}{2\sqrt{\pi}\xi^{1/4}(2\epsilon)^{1/3}}\exp\left[\sigma-\frac{2}{3}\xi^{\frac{3}{2}}\right].
\end{equation}
Let us first calculate the term appearing in the exponent. Since
\begin{gather}
		b_{1}^{3}+b_{2}^{3}=2b^{3} \\
		(b_{1}b_{2})^{3/2}=b^{3}\sqrt{1-\epsilon^{2}/b^{6}} \\
		b_{1}^{3}c_{1}-b_{2}^{3}c_{2}=b^{3}(c_{1}-c_{2})+(c_{1}+c_{2})\epsilon,
\end{gather}
we have
\begin{equation}
		\sigma=\frac{a^{3}b^{3}}{6\epsilon^{2}}-\frac{ab^{3}(c_{1}-c_{2})}{2\epsilon}-\frac{a(c_{1}+c_{2})}{2}
\end{equation}
and
\begin{align}
		\xi^{\frac{3}{2}}&=\frac{a^{3}(b_{1}b_{2})^{3/2}}{4\epsilon^{2}}\left(1-\frac{2(c_{1}-c_{2})\epsilon}{a^{2}}\right)^{3/2} \nonumber \\
		&=\frac{a^{3}b^{3}}{4\epsilon^{2}}\left[\left(1-\frac{\epsilon^{2}}{b^{6}}\right)\left(1-\frac{2(c_{1}-c_{2})\epsilon}{a^{2}}\right)^{3}\right]^{1/2} \nonumber \\
		&=\frac{a^{3}b^{3}}{4\epsilon^{2}}\left[1-\frac{3(c_{1}-c_{2})\epsilon}{a^{2}}
		+\left(\frac{3(c_{1}-c_{2})^{2}}{2a^{4}}-\frac{1}{2b^{6}}\right)\epsilon^{2}\right] \nonumber \\
		&+O(\epsilon).
\end{align}
We get
\begin{equation}
		\sigma-\frac{2}{3}\xi^{\frac{3}{2}}
		=-\frac{a(c_{1}+c_{2})}{2}-\frac{b^{3}(c_{1}-c_{2})^{2}}{4a}+\frac{a^{3}}{12b^{3}}+O(\epsilon).
\end{equation}
Next, we calculate the factor in front of the exponent. Since
\begin{equation}
		\xi^{1/4}=\frac{\sqrt{a}(b_{1}b_{2})^{1/4}}{(2\epsilon)^{1/3}}\left(1-\frac{2(c_{1}-c_{2})\epsilon}{a^{2}}\right)^{1/4},
\end{equation}
\begin{widetext}
we have 
\begin{align}
		\frac{1}{2\sqrt{\pi}\xi^{1/4}(2\epsilon)^{1/3}}&=
		\frac{1}{2\sqrt{a\pi b}}\left[\left(1-\frac{\epsilon^{2}}{b^{6}}\right)\left(1-\frac{2(c_{1}-c_{2})\epsilon}{a^{2}}\right)^{3}\right]^{-\frac{1}{12}} \nonumber \\
		&=\frac{1}{2\sqrt{ab\pi}}\left[1+\frac{(c_{1}-c_{2})\epsilon}{2a^{2}}-\left(\frac{7(c_{1}-c_{2})^{2}}{8a^{4}}-\frac{1}{12b^{6}}\right)\epsilon^{2}\right] 
		+O(\epsilon^{3}).
\end{align}
At last, we have
\begin{equation}
		I(\epsilon)=
		\frac{1}{2\sqrt{ab\pi}}\left[1+\frac{(c_{1}-c_{2})\epsilon}{2a^{2}}-\left(\frac{7(c_{1}-c_{2})^{2}}{8a^{4}}
		-\frac{1}{12b^{6}}\right)\epsilon^{2}+O(\epsilon^{3})\right] 
		\exp\left[\frac{a^{3}}{12b^{3}}-\frac{a(c_{1}+c_{2})}{2}-\frac{b^{3}(c_{1}-c_{2})^{2}}{4a}+O(\epsilon)\right]	
\end{equation}
and the leading term in the asymptotic limit is
\begin{equation}
		\lim_{\epsilon\rightarrow0^{+}}I(\epsilon)=\frac{1}{2\sqrt{ab\pi}}\exp\left[\frac{a^{3}b^{3}}{72}-\frac{a(c_{1}+c_{2})}{2}-\frac{b^{3}(c_{1}-c_{2})^{2}}{4a}\right].
\end{equation}
\end{widetext}

\end{document}